\input harvmac
\sequentialequations
\def\Title#1#2#3{#3\hfill\break \vskip -0.35in
\rightline{#1}\ifx\answ\bigans\nopagenumbers\pageno0\vskip.2in
\else\pageno1\vskip.2in\fi \centerline{\titlefont #2}\vskip .1in}


\def\ket#1{| #1\rangle}

\def\half{{1\over 2}}

\def\R{\hbox{\rm I \kern-5pt R}}

\font\ticp=cmcsc10
\def\ajou#1&#2(#3){\ \sl#1\bf#2\rm(19#3)}
%
%
\lref\griff{Griffiths,R., \ajou J. Stat. Phys. &36 (84) 219.}

\lref\griffchqr{Griffiths,R., \ajou Phys. Rev. A &54 (96) 2759.} 

\lref\omnes{Omn\`es,R., \ajou J. Stat. Phys. &53 (88) 893.}

\lref\omnesbook{Omn\`es,R., ``The Interpretation of Quantum
Mechanics'' (Princeton University Press, Princeton 1994).}

\lref\gmhsantafe{Gell-Mann,M.~and Hartle,J., ``Complexity, Entropy,
and the Physics of Information, SFI Studies in the Sciences of
Complexity, Vol. VIII,'' (Edited by W.~Zurek) 
(Addison Wesley, Reading 1990), p.~425. }

\lref\grw{Ghirardi, G., Rimini, A.~and Weber,T., \ajou Phys. Rev. D &34
(86) 470.}

\lref\gpr{Ghirardi, G., Pearle, P.~and Rimini,A., \ajou Phys. Rev. A &42
(90) 78.}

\lref\ggr{Ghirardi, G., Grassi,R.~and Rimini,A., \ajou Phys. Rev. A &42
(90) 1057.}

\lref\gisin{Gisin,N., \ajou Helv. Phys. Acta. &62 (89) 363.} 

\lref\hartleone{Hartle,J., ``Quantum Cosmology and Baby
Universes, Proceedings of the 1989 Jerusalem Winter School on
Theoretical Physics'' 
(Edited by Coleman,S., Hartle,J., Piran,T. and Weinberg,S.)
(World Scientific, Singapore, 1991), p.~65.}

\lref\rudolphone{Rudolph,O., \ajou Int. J. Theor. Phys. &35 (96)
  1581.}

\lref\rudolphtwo{Rudolph,O., \ajou J. Math. Phys. &37 (96) 5368.}

\lref\hartlesqm{Hartle,J., \ajou Phys. Rev. D &10 (91) 3173.}

\lref\ishamql{Isham,C., \ajou J. Math. Phys. &23 (94) 2157.}

\lref\ILqtl{Isham,C.~and Linden,N., \ajou J. Math. Phys. &35 (94) 5452.}

\lref\ILSclass{Isham,C., Linden,N.~and Schreckenberg,S., 
\ajou J. Math. Phys. &35 (94) 6360.}

\lref\gmhstrong{Gell-Mann,M.~and Hartle,J., gr-qc/9509054.}

\lref\akcontrary{Kent,A., \ajou Phys. Rev. Lett. &78 (97) 2874.}

\lref\akordered{Kent,A., gr-qc/9607073.}

\lref\goldsteinpage{Goldstein,S.~and Page,D., \ajou Phys. Rev. Lett. 
&74 (95) 3715.}

\lref\pitowskyhemmo{Pitowksky,I.~and Hemmo,M., unpublished.}

\lref\dowkerkentone{Dowker,F.~and Kent,A.,
\ajou J. Stat. Phys. &82 (96) 1575.}

\lref\dowkerkenttwo{Dowker,F.~and Kent,A., \ajou Phys. Rev. Lett. &75
(95) 3038.}

\lref\omnesreview{Omn\`es,R., \ajou Rev.~Mod.~Phys. &64 (92) 339.}

\lref\ishamtopos{Isham,C., \ajou Int. J. Theor. Phys. &36 (97) 785.}

\lref\griffchoice{Griffiths,R., quant-ph/9708028.}

\lref\aktwo{Kent,A., \ajou Phys. Rev. A &54 (96) 4670.}

\lref\kentmcelwaine{Kent,A.~and McElwaine,J., \ajou Phys. Rev. A &55
(97) 1703.}

\lref\ishamlindeninfoentropy{Isham,C.~and Linden,N., \ajou Phys. Rev. A &55
(97) 4030.}

\lref\gmhprd{Gell-Mann,M.~and Hartle,J., \ajou Phys. Rev. D
&47 (93) 3345.}

\lref\bellgrw{Bell,J.~, ``Speakable and Unspeakable in Quantum 
Mechanics'' (Cambridge University Press, Cambridge 1988), p.~201.}

\lref\diosi{Di\'osi,L.~, \ajou Phys. Rev. A &40 (89) 1165.}

\lref\pearle{Pearle,P.~, ``Quantum Chaos -- Quantum Measurement''
(Edited by Cvitanovic, P., Percival, I.~and Wirzba,A.) 
(Kluwer Academic Publishers, Dordrecht 1992), p. 283.} 

\lref\percival{Percival,I.~, \ajou Proc. Roy. Soc. Lond. A &447 (94) 189.}

\lref\pearlesquires{Pearle,P.~and Squires,E., \ajou Phys. 
Rev. Lett. &73 (94) 1.}

\Title{
\vbox{\baselineskip12pt\hbox{ DAMTP-1997-117}\hbox{ gr-qc/9809026}{}}
}{\centerline{Quantum Histories}}{~}
\centerline{{\ticp Adrian Kent}}
\vskip.1in
\centerline{\sl Department of Applied Mathematics and
Theoretical Physics,}
\centerline{\sl University of Cambridge,}
\centerline{\sl Silver Street, Cambridge CB3 9EW, U.K.}

\bigskip

\centerline{\bf Abstract}
{There are good motivations for considering some 
type of quantum histories formalism. 
Several possible formalisms are known, defined by 
different definitions of event and by different selection 
criteria for sets of histories.  
These formalisms have a natural interpretation, according
to which nature somehow chooses one set of histories from
among those allowed, and then randomly chooses to realise
one history from that set; other interpretations are 
possible, but their scientific implications are 
essentially the same.  

The selection criteria proposed to date are reasonably natural,
and certainly raise new questions.  
For example, the validity of ordering inferences which we normally
take for granted --- such as that a particle in one region is 
necessarily in a larger region containing it --- depends on whether 
or not our history respects the criterion of ordered
consistency, or merely consistency.   

However, the known selection criteria, including consistency and
medium decoherence, are very weak.  It is not possible to 
derive the predictions of classical mechanics or Copenhagen
quantum mechanics from the theories they define, even 
given observational data in an extended time interval. 
Attempts to refine the consistent histories approach so 
as to solve this problem by finding a definition of 
quasiclassicality have so far not succeeded. 

On the other hand, it is shown that dynamical collapse models, 
of the type originally proposed by Ghirardi-Rimini-Weber,
can be re-interpreted as set selection criteria within 
a quantum histories framework, in which context they 
appear as candidate solutions to the set selection problem. 
This suggests a new route to relativistic generalisation 
of these models, since covariant definitions of a quantum event 
are known. 
\medskip\noindent
\medskip
Contribution to Proceedings of the 104th Nobel Symposium,  ``Modern 
Studies of Basic Quantum Concepts and Phenomena'', Gimo, June 1997; 
to appear in Physica Scripta (1998).  
\eject

\newsec{Introduction} 

The orthodox view of quantum theory has come under attack 
from several quarters lately, rather to the mystification 
of those who think the existing theory perfectly adequate.    
In particular, many can see no scientific motivation for a ``consistent 
histories'' interpretation of quantum theory\refs{\griff, \griffchqr, 
\omnes, \omnesbook, \gmhsantafe} which
apparently cannot be experimentally refuted without also 
refuting Copenhagen quantum theory. 
And, while the dynamical collapse models proposed by 
Ghirardi-Rimini-Weber\refs{\grw} and 
others\refs{e.g.\  \gpr,\ggr,\gisin} at least offer testable
alternatives to quantum theory, their broader scientific 
motivations are also not widely understood. 

I will try here to motivate a histories approach to quantum
theory, to describe the problems (which are serious) with the existing
consistent histories proposals, and to give a unifying picture 
in which dynamical collapse
models can be re-interpreted within a history framework and seen
as attempts to address the problems afflicting the existing 
proposals.  
This is not meant as a survey, but rather
a personal view of some of the key ideas in
the field; very different views of the consistent histories 
approach can be found among the papers cited. 
I am particularly indebted to a collaboration
with Dowker,\refs{\dowkerkentone,\dowkerkenttwo} from 
which some of the key points made below derive. 

The existing consistent histories formulations of quantum 
theory have so far proved to be of little or no direct
scientific use {\it per se}. 
But it seems to me that some of the motivations 
advanced by consistent historians are nonetheless valid and 
some of the questions they raise are scientifically interesting.
The approach, I will argue, should be seen as an incomplete 
but interesting research program --- a program that has run into 
serious problems, and whose basic assumptions now need
to be reconsidered, but one that includes some natural ideas
which are worth pursuing.  
Its ultimate aim, I will argue, must be to solve the so-called 
set selection problem: that is, to define a histories 
formulation of quantum theory from which successful existing
theories --- in particular, classical mechanics and the Copenhagen
interpretation of quantum theory --- can be derived, within
their domains of validity. 

It is hard to see how to produce a quantum histories
formulation that is both simple and in precise agreement
with standard quantum theory.  But the derivation of 
existing theories requires only agreement with known experiment, 
not necessarily precise agreement with standard quantum theory. 
It will be shown that dynamical collapse models belong to the  
class of generalised quantum histories theories.
Viewed in this way, they illustrate  
how the set selection problem can be satisfactorily solved 
in the non-relativistic limit.  

This suggests another reason for taking collapse models
seriously: in the non-relativistic case,
they are the best solutions known to the 
problem of defining a quantum histories theory that 
satisfies fairly minimal scientific criteria.  
It seems to me also to suggest that it is more natural, and 
more likely to be fruitful, to look for relativistic 
generalisations of collapse models in the framework of 
history models than to try to find some form of 
relativistic generalisation of the stochastic 
differential equations that define the existing models.  

\newsec{The case for quantum histories} 

It can be hard to look afresh at so worked-over a topic as the
scientific status of quantum theory.  
Maybe it is helpful to recall an earlier scientific 
debate --- that over the behavourist program 
in experimental psychology. 

Like many of the founders and developers of quantum theory, 
radical behaviourists --- most notably, Skinner --- took a rigorously   
instrumentalist view.  They saw the proper task 
of psychology as the generation of theories predicting responses
to stimuli, without invoking intermediate
explanatory hypotheses about the mental states 
of the subject tested.  Such hypotheses were, in their view, 
scientifically meaningless, referring as they do to unmeasurable 
quantities.  The common language of mental 
states was to be understood, at best, as a sort of improper 
shorthand for statements about earlier stimuli.  ``The subject is angry'', 
for example, might perhaps more accurately be translated as
``the subject has been repeatedly prodded with a stick''.  
The mind, in other words, was a black box --- 
to be prepared for experiment by stimuli, to be 
investigated through its responses to further stimuli, and to be 
described theoretically by the correlations between the responses 
and stimuli.  

Skinnerian behaviourism was never an unquestioned orthodoxy, and has
now been almost completely abandoned.  
It was not rejected because of internal inconsistency, or because
psychologists rejected instrumentalism on purely philosophical
grounds. It simply became increasingly apparent that its 
axioms were obstructing good science.  
Mind-states are, if nothing else, useful theoretical constructs, 
not always, in practice, fully explicable in terms of prior 
stimuli.  (Think of depression, for example.) 
And learning turns out to be a subtler process than Skinnerian  
accounts allow.  As Chomsky showed, the acquisition of language
cannot be explained purely as a product of classical stimulus-response
conditioning: no scientific explanation of the fact that we speak  
grammatically can proceed from Skinnerian axioms alone.  

The Copenhagen view of a quantum system very much resembles
the Skinnerian view of the mind: statements about the system's
behaviour between preparation and measurement are illegitimate.   
Consistent historians, like many other critics of quantum orthodoxy, 
reject this, taking seriously the idea that quantum 
events take place, whether or not anyone is looking.  
If we assume that the events somehow change the dynamics, clearly 
we have something to test. 
But let us suppose for the moment, as consistent historians do, that
unobserved quantum events take place, but that the dynamics are
unaltered.  Could this sort of interpretation still lead to 
interesting new science?  

Probably the best case that it could comes from cosmology.  
When cosmological ideas are discussed,
we generally proceed from a quantum description of the 
initial or very early conditions, and then 
explain its consequences in terms of successive events and 
processes which, it is to be hoped, together explain the present
state of the cosmos.
Almost everyone thinks this way;
almost everyone realises it is illegitimate.
(Successive events?  During the evolution of the closed quantum 
system that defines the universe?  In the pre-classical era?)
What can we really mean?  

One possible tactic is to try to interpret all hypotheses about 
past events in terms of present observations.  
But will this always be possible?  
We cannot hope to 
calculate the probabilities of present states
directly from the initial conditions in a realistic theory: 
any successful cosmological theory is bound to involve a 
long chain of reasoning involving many successive events.
Practically speaking, will the present consequences of every
intermediate hypothesis be calculable?  And in principle, is it 
clear that good theories will involve 
only quantum events which {\it have} directly and independently
observable present consequences?  
Any more, say, than it is clear that mind-states can be reduced 
to stimuli, in practice or in principle?

Different people have different intuitions on these questions 
at present.  Time will tell: for what it is worth, I follow
Hartle\refs{\hartleone} in believing that the reduction 
will probably be impossible. 
Clearly, if future cosmological theories turn out to be  
irreducible to present observations, and if no more radical 
modification of quantum theory takes place, some sort of quantum
histories formalism will be needed.  
We will need to be able to make sense of the notion that
some sequence of events, drawn from some larger set of 
possibilities, took place during the evolution of a closed
system.  How then might the possible events and histories 
be defined? 

\newsec{Events and histories} 

A quantum event ought, presumably, to have a mathematical
representation that fits naturally into a standard 
formulation of quantum theory, in a way that allows us 
to consider histories: collections of events occurring 
during the evolution of a system.
We need, too, a natural rule for defining sample spaces of 
possible histories, with a probability measure.  
And it must be possible, at least in principle, to represent 
at least some familiar physical events --- the results 
of measurements, for example --- in terms of the defined 
histories, in order to connect the definitions with physics 
as we know it.  

Several different definitions satisfying these demands have been
proposed.  
The simplest version of quantum event
is defined by fixing a time, $t$, at which it takes place, 
together with a Heisenberg picture 
projection $P$, which --- so to speak --- 
says what happened: the system was in the range of $P$ at time $t$. 

A complete list of exclusive alternative events at time $t_j$
is then given by a projective decomposition of the identity:
\eqn\events{\sigma_j = \{ P_j^{(1)} , \ldots , P_j^{(n_j )} \} \, ; \qquad
\sum_{i=1}^{n_j} P_j^{(i)} = I \, ; 
\qquad P_j^{(i)} P_j^{(i')} = \delta^{i i'} P_j^{(i')} \, . }
An elementary history $H$ is a list of projections $\{ P_1^{(i_1 )} , \ldots , 
P_n^{(i_n )} \}$ 
at distinct times $t_1$ to $t_n$, and a complete set of 
exclusive alternative histories is defined by all the possible 
combinations of projections from any fixed set of 
projective decompositions at distinct times 
\eqn\set{S = \{ \sigma_1 , t_1 ; \ldots ; \sigma_n , t_n \} \, .}
The set $S$ defines a sample space, and a probability distribution
is defined by defining the probability of an
elementary history:  
\eqn\prob{
P (H) =  \Tr ( P^{(i_n )}_n \ldots P^{(i_1 )}_1 \rho P^{(i_1 )}_1
\ldots P^{(i_n )}_n ) \, ,}
where $\rho$ is the initial density matrix of the system at $t=0$. 
Note that these quantities satisfy the probability axioms --- i.e. they
are non-negative and add to one --- without any further restriction
on the set $S$. 

This definition can usefully be generalised to ``unsharp'' 
events represented by positive operators, an idea first 
investigated by Rudolph.\refs{\rudolphone,\rudolphtwo}  
For the discussion here the
following simple definitions (not equivalent to Rudolph's) will
be adequate.  An unsharp event, again at fixed time $t$, is defined 
by any positive operator $A$.  
A complete list of exclusive alternative
unsharp events at time $t_j$ is given by a decomposition of the 
identity into distinct positive operators:
\eqn\povdecomp{
\sigma_j = \{ A_j^{(1)} , \ldots , A_j^{(n_j )} \} \, ; \qquad
\sum_{i=1}^{n_j} A_j^{(i)} = I
\, ;  \qquad A_j^{(i)} \neq A_j^{(i')} {\rm~for~} i \neq i' \, .}

Elementary histories and complete sets of exclusive alternative
histories are defined by generalising the projection operator
definitions in the obvious way, and the probability of an 
elementary history $H = \{ A_1^{i_1} , \ldots , 
A_n^{i_n} \}$ is given by
\eqn\povprob{
P (H) =  \Tr ( (A^{(i_n )}_n )^{\half} \ldots (A^{(i_1 )}_1 )^{\half} \rho
( A^{(i_1 )}_1 )^{\half} \ldots (A^{(i_n )}_n )^{\half} ) \, .}
Again, these quantities automatically behave as probabilities. 

Alternatively, events can be defined by partitions
of the configuration space path integral.\refs{\hartlesqm,\hartleone}  
Temporarily moving to the Schr\"odinger picture and 
taking the initial state $\ket{\psi}$ to be pure, 
we can use a partition of the space of paths 
$\{ c_{\alpha} \}_{\alpha \in A}$ to define branches
$\ket{\psi_{\alpha}}$
and class operators $C_{\alpha}$ by 
\eqn\class{
\ket{ \psi_\alpha } \equiv C_\alpha \ket{ \psi } =
\int\nolimits_{c_\alpha} \delta q
\ \exp \bigl(iS[q(\tau)]/\hbar\bigr) \ket{\psi} \, .}
Summing over all paths gives the usual evolution, so that 
\eqn\classsum{
\sum\nolimits_{\alpha} C_{\alpha} = e^{-iHT/\hbar}\ .}

The probability weights
\eqn\classprob{ P (c_{\alpha} ) = 
\big\Arrowvert  \big| \psi_\alpha \bigr\rangle 
\big\Arrowvert^2\ \,}
can --- provided 
$ \sum\nolimits_{\alpha} p( c_{\alpha} ) $ is finite 
--- be normalised to probabilities for the elementary 
events $c_{\alpha}$. 

From the fundamental point of view, this is perhaps the most 
interesting approach.  Not only does it allow events to be 
defined covariantly in a background spacetime --- the classes 
$c_{\alpha}$ could correspond to paths crossing or not 
crossing various space-time regions, for example --- 
but it can also be extended, at least formally, 
to sum-over-manifold formulations of quantum gravity.  

It is worth mentioning for completeness (though they will
not be needed here) that natural generalisations of 
projection-valued events corresponding to multi-time 
propositions can also be defined --- a simple example
being that the composite event defined by 
the conjunction of two elementary events can 
be represented by the tensor product 
of the relevant projection operators.  
One of the aims of the Isham-Linden-Schreckenberg
version\refs{\ishamql,\ILqtl,\ILSclass} of consistent 
histories is to investigate general definitions of multi-time 
events and their physical relevance.  

Logically, all of these definitions make perfect sense.  
Scientifically, they have two serious problems. 
First, whichever definition is used, probability distributions
are defined on uncountably many different sets of histories, and 
it is not clear which of these sets (if indeed any) are 
appropriate in any given physical situation.  
Second, though the distributions obviously satisfy the mathematical
axioms for probabilities, it is not at all clear that they 
are physically meaningful.  

\newsec{``Consistency'' and other selection criteria}

The consistent histories approach attempts to address the 
last-mentioned problem, by restricting to sets of histories  
on which the probability distributions have properties  
which are, at least arguably, desirable.
It is not absolutely clear, though, that these properties are 
always necessary for physical relevance.  And it is clear that
they are not sufficient: the approach does
not address the first problem, as we will see. 

So it should perhaps be stressed at the outset that 
none of the selection criteria discussed here has any
privileged status.  There is no logical or mathematical
requirement to impose any criterion --- the technical
term ``consistency'' is here rather misleading.  
The various criteria proposed to date are simply guesses.  
Even if the basic idea of a quantum histories approach
is correct, so that for any given physical system there is some 
identifiable set of histories which probabilistically predicts 
the past and future events from initial data, it is  
possible that this set satisfies none of these criteria.  

That said, one has to start somewhere, and the various criteria 
do characterise interesting properties. 
So far as is known, not (quite) every criterion considered can 
be naturally extended to cover every notion of a history, but 
the projection operator notion of events illustrates
the full spectrum of possibilities. 
It is simplest again to consider non-relativistic quantum
theory, in the Heisenberg picture. 

Griffiths' original consistency criterion for a set $S$ of histories
is that the probability formula ought to respect the 
rule that the union of events can be represented by summing 
the corresponding projection operators, so that 
\eqn\sumrules{\Tr ( Q_n \ldots Q_1 \rho Q_1 \ldots Q_n ) 
=  \sum_{i_1 \in I_1 \ldots i_n \in I_n } \Tr ( P^{(i_n )}_n \ldots
P^{(i_1 )}_1 \rho P^{(i_1 )}_1 \ldots P^{(i_n )}_n ) \, ,}
for all projections 
\eqn\qs{
Q_j = \sum_{i_j \in I_j} P^{(i_j )}_j }
given by sums of the elementary projections at time $t_j$ in $S$.  
This holds\refs{\griff} if and only if
\eqn\decohgriff{ {\rm Re}\,
(\Tr ( P^{(i_n )}_n \ldots P^{(i_r )}_r \ldots P^{(i_1 )}_1
                               \rho P^{(i_1 )}_1
\ldots P^{(i'_r )}_r \ldots P^{(i_n )}_n  )) =
\delta_{i_r i'_r } p(i_1 \ldots i_n ) \, ,}
for all $r$ and all choices of $i_1 , \ldots , i_n$ and $i'_r$, 
where $p(i_1 , \ldots , i_n )$ is shorthand for the history
probability \prob\ .

A consistent set of histories defined by positive operator events
can also naturally be defined by extending equation \sumrules\ , 
so that we say a set 
$S = \{ \sigma_1 , t_1 ; \ldots ; \sigma_n , t_n \}$ defined
by positive operator decompositions of the form \povdecomp\ 
is consistent if 
\eqn\sumrulespov{\Tr ( B_n \ldots B_1 \rho B_1 \ldots B_n ) 
=  \sum_{i_1 \in I_1 \ldots i_n \in I_n } \Tr ( B^{(i_n )}_n \ldots
B^{(i_1 )}_1 \rho B^{(i_1 )}_1 \ldots B^{(i_n )}_n ) \, ,}
where 
\eqn\bsqrt{ B^{(i)}_j = (A^{(i)}_j )^{\half} \, ; \qquad 
B_j = ( \sum_{i \in I_j} A^{(i)}_j )^{\half} \, . }

The main point of these definitions is that the probability for an
individual event in a history belonging to a consistent set
can be calculated very simply even when one is partially
or completely ignorant of past events.
For example, equation \sumrules\ implies that, if nothing is known
about the past, the probability of $P^{i_n}_n$ would simply be 
\eqn\probpinn{ Tr  ( P^{(i_n )}_n  \rho P^{(i_n )}_n ) \, ;}
equation \sumrulespov\ implies the analogous statement in 
the case of positive operator events.   

The stronger condition of {\it medium decoherence}\refs{\gmhsantafe}
requires that
\eqn\decohgmh{ \Tr ( P^{(i_n )}_n \ldots P^{(i_1 )}_1
                               \rho P^{(j_1 )}_1
\ldots P^{(j_n )}_n ) =
\delta_{i_1 j_1 } \ldots \delta_{i_n j_n } p(i_1 \ldots i_n ) \, .}
Again, the terminology can mislead the unwary:
medium decoherence is mathematically a natural condition, 
but it does not generally identify sets of histories describing events 
characterised by decoherence in the ordinary physical sense.   
Gell-Mann and Hartle have also investigated\refs{\gmhstrong} a 
criterion of {\it strong decoherence}.  
This proposal, however, must be considered exploratory: as it 
stands, every medium decoherent set is strongly decoherent.   

These early attempts at selection criteria are very weak, in 
a sense to be made more precise in the next sections.  One reflection
of that weakness is the disconcerting fact that it is easy to 
find examples in which they allow two or more contrary 
propositions --- statements corresponding to orthogonal 
projections --- to be retrodicted from the same data, each with 
probability one, in different sets.\refs{\akcontrary, \akordered}  
For example, one can arrange so that
different sets imply that, at a given time, a given particle 
was in one of two different (non-intersecting) boxes. 
A consequence of this is that logical 
implications which we normally take for granted are violated.
For example, according to a consistent or decoherent 
histories analysis, it can lead to a logical contradiction 
to infer from the observation that a particle was in a given 
region at a given time that the particle was 
in a larger region containing the first.  

A somewhat stronger criterion, which eliminates this problem, can be 
defined as follows.\refs{\akordered} 
The standard partial ordering on projections, according to which
$P \leq Q$ if and only if $PQ = QP = P$, defines 
a natural partial ordering on the class of histories:
\eqn\ordering{
\{ P_1 , t_1 ; P_2 , t_2 ; \ldots ; P_n , t_n \} \leq 
\{ Q_1 , t_1 ; Q_2 , t_2 ; \ldots ; Q_n , t_n \} 
\Leftrightarrow P_i \leq Q_i {\rm~for~all~}i \, .}
Histories differing only by the inclusion of copies of the 
identity operator at various times are here regarded as equivalent.
We now define an {\it ordered consistent history} to be a history $H$,
belonging to some medium decoherent set $S$, with the property
that if $H'$ is a history belonging to any other medium
decoherent set $S'$ such that $H \leq H'$ then we have 
$P(H) \leq P(H')$, and similarly $H \geq H'$ implies
$P(H) \geq P(H')$. 
An {\it ordered consistent} set is then a set all of whose 
elementary histories are ordered consistent.  
Ordered consistency can be defined similarly for the positive
operator and path integral partition definitions of a quantum event.  

There is some room for doubt as to whether 
ordered consistency is {\it too} strong a criterion: it has not 
been convincingly demonstrated that all familiar physics can 
necessarily be described by ordered consistent sets of 
histories.  It would be particularly good to resolve this
question, since either answer leads to an interesting 
conclusion.  If arguments can be found that ordered consistent
sets are adequate, then ordered consistency defines the
strongest and least problematic quantum histories approach
currently available that respects standard quantum dynamics.  
Conversely, if ordered consistent
sets can be shown to be inadequate, then standard ordering 
implications would have to be abandoned,
with radical implications for our scientific worldview: 
it would no longer be possible to infer that a measurement of any 
observable in any range implies that it lay in any strictly
larger range, for example.\refs{\akordered}

Two other criteria --- linear positivity\refs{\goldsteinpage} and
feasibility\refs{\pitowskyhemmo} --- have also recently been 
defined.  Both are weaker than consistency: for them to  
be of independent use in solving the problems considered here, 
some plausibly physically relevant refinement incompatible 
with consistency would have to be found.

To summarise, we have a spectrum of reasonably natural criteria
which, as it happens, can be ordered in terms of increasing
refinement: feasibility, linear positivity, consistency, 
medium decoherence, ordered consistency.  
There are thus at least six candidate quantum histories schemes, based
on unrestricted sets of histories or
on sets selected by one of the five criteria, 
and most of these schemes can be defined for each of the 
three natural notions of quantum event discussed to date.

Fortunately, these schemes all share some key features, 
which means that in assessing their present scientific status
they need not all be discussed separately. 
Unfortunately, as we will see, this is largely because 
all the known criteria are far too weak.  

\newsec{Interpreting history-based schemes} 

Broadly speaking, there are two views of what the consistent
histories formalism, or any other new version of quantum theory,
could be good for.  
According to one, the idea is to understand what 
quantum theory really means, in some abstract idealistic sense. 
According to the other, the 
ultimate aim is to make scientific progress in the more concrete
sense of generating new testable theories, allowing new calculations, 
and making new predictions, while retaining the successes of the 
Copenhagen interpretation --- in short to go beyond
Copenhagen quantum theory in something like the way that general
relativity goes beyond Newtonian gravity.  
Part of the reason why the subject is so controversial, I suspect, 
is that it is sometimes the battleground for a kind of undeclared guerilla
conflict between these motives, which perhaps are not always 
cleanly disentangled even in authors' own minds.  

I would place recent attempts by 
Griffiths,\refs{\griffchqr} Omn\`es\refs{\omnesreview, \omnesbook} ---
note, incidentally, that Omn\`es' theory of ``truth''\refs{\omnesreview} 
is almost entirely wrong,\refs{\dowkerkentone} as Omn\`es now accepts ---  
and Isham\refs{\ishamtopos} to set out logical structures for the 
consistent histories formalism in the first camp.  
It seems to me these ideas can only be 
appraised on their own terms: at the moment they promise 
no new concrete scientific yield. 

In practical terms, however, all the interpretational ideas which 
have been set out for the consistent histories approach have the 
same scientific implications, with one minor caveat that I will 
address in a moment.  The following discussion applies equally
to all the other quantum histories approaches.

Everyone agrees that the  
generally incompatible pictures of physics given by the 
uncountably many different consistent sets have to be 
assigned equal fundamental status.  The formalism does 
not distinguish amongst them: to do that would need 
further selection criteria, which would define a different 
quantum histories approach. 
However, the physics we actually see is described
by just one history. 

There are three slightly different ways of interpreting the
situation.  The most economical is (i) that nature has 
chosen, somehow, one consistent set --- since it is not known 
if there is any natural measure on the full class of 
consistent sets, we cannot be more precise --- which 
defines the sample space of histories and their probabilities.  
Nature then randomly chooses, according to these probabilities, 
to realise one history, which must turn out to be the one we see. 

One could say, alternatively, (ii) that one history is randomly chosen from
every consistent set, or (iii) that all the histories from every consistent
set are realised in numbers proportional to their probabilities.   
In either case, we must somehow find ourselves attached to precisely
one of the realised histories. 
A possible attraction of these last two ways of putting things, one might
think, is that they allow the possibility that the type of 
history we find ourselves in is not determined randomly, nor 
by new fundamental selection criteria, but by something to do 
with us --- specifically, that our consciousness somehow 
attaches itself to quasiclassical histories.  
Some such hypothesis, within the second picture, seems indeed 
to underlie some of Gell-Mann and Hartle's and 
Griffiths' arguments,\refs{\gmhsantafe, \griffchoice , \dowkerkentone}
though for obvious reasons it has not been fleshed out.    
However, even if these ideas could be made concrete, there 
is a compelling argument to show that they would not work, 
essentially because any given quasiclassical history
belongs to many inequivalent consistent sets.
This makes it impossible, in the second picture, to derive 
the predictions of classical mechanics or Copenhagen quantum 
mechanics, even under the assumption
that we will persistently experience quasiclassicality.\refs{\aktwo} 
(The third picture, I believe, suffers from a similar problem, 
though no discussion has appeared in print.) 

Given this failing, and since the pictures are 
equivalent unless some unknown theory of consciousness is 
attached, we need only consider the first picture. 
Nature, it says, is described by one of 
a large number of sub-theories, which correspond to the 
various sets of histories --- in rather the same sort of 
way, for example, as general relativity says that nature
is described by one of the solutions of Einstein's 
equations.  The sub-theories in the consistent histories formalism 
(and the other quantum histories formalisms) are probabilistic
rather than deterministic, of course, since choosing the
set only determines the space of possible histories.  
But that itself is no drawback (except to diehard determinists).  
The key question is what we can achieve with
this collection of sub-theories.  How far can the 
analogy be pressed?

\newsec{Why the known criteria are too weak}  

General relativity is almost universally seen as the paradigm of 
a successful physical theory, incorporating and unifying as it
does special relativity, Newtonian gravity, and classical
mechanics.  Of course, its incompatibility with quantum theory 
and its singularities suggest that it will eventually be supplanted.  
But setting aside these problems, the theory 
has what might be, but usually is not, seen as an intrinsic 
weakness: it does not tell us {\it which} solution of
Einstein's equations nature has chosen. 
This is not seen as a significant weakness 
since Einstein's equations can be solved locally 
given initial data on a hypersurface, which in turn can be
approximated by carrying out measurements in a local region.  
We thus can and do carry out observations to determine which
local solution is relevant, and hence make predictions 
and retrodictions within general relativity.  In particular, 
in this way, we can derive the predictions of Newtonian gravity and 
classical mechanics within their domain of validity, which 
we understand to be the weak field limit of general relativity.   
In short, we understand when and why Newtonian gravity and
classical mechanics hold true, and how to tell whether they
will hold true in any given physical situation.  

An analogously successful quantum histories approach would 
incorporate classical mechanics and Copenhagen quantum
mechanics in a similar way.  It need not provide a 
theory of the quantum boundary conditions --- we can
assume for the sake of the argument that these are 
fixed.  Nor need it supply {\it a priori} the set 
from which nature chooses the realised history.    
But, applied to non-relativistic quantum mechanics, it should 
explain how to identify that set {\it post hoc}, 
on the basis of observations within some finite time interval. 
(In the relativistic case, it should presumably explain how to extrapolate
a local description of the set, given observations in some
finite space-time region.) 
And it must characterise the domain of validity of classical mechanics
and Copenhagen quantum theory and explain what types of observations
are necessary in order to infer predictions and retrodictions 
within those theories. 

No quantum histories approach defined by any of the existing
criteria satisfies any of these demands --- quite the reverse.
For example, in any physically reasonable model, it is 
impossible to identify the correct medium decoherent set, or infer  
any of the decompositions defining its past or future events, 
on the basis of any set of observations taking place in any
finite time interval.\refs{\dowkerkentone, \dowkerkenttwo} 
If we know the initial density matrix $\rho$ and the 
hamiltonian, and we observe that the series of events 
defined by projections $P_1 , \ldots , P_n$ took 
place at times $t_1 < \ldots < t_n$, we 
still generally cannot identify any of the projective 
decompositions which define the set, from which this partial
history is drawn, at times before $t_1$ or after $t_n$: 
there are almost always many incompatible medium decoherent 
sets which incorporate the observed data
and make incompatible retrodictions of the past and
predictions of the future.

In short, almost nothing can be unambiguously predicted or 
retrodicted on the basis of the medium decoherent histories formalism
alone.  We {\it can} make statements of the form ``{\it if} the 
relevant medium decoherent set is $S$, then the following future 
(or past) events are possible (or may have occurred), with the 
following probabilities''.  But we cannot identify $S$, and 
without doing so we cannot derive classical mechanics or Copenhagen
quantum mechanics, or fully explain their successes.  
This is the main reason why, it seems to me, the existing
quantum histories formalisms can only be viewed as part of
a seriously incomplete research program.  
What seems to be required, if it is to be completed, is a 
criterion sufficiently strong that data in a finite time 
interval can either positively identify the relevant set of 
quantum histories or at least constrain the range of possibilities 
sufficiently that standard physics can be derived. 
This is the so-called {\it set selection 
problem}.\refs{\dowkerkentone}  
A set selection criterion need not, of course, necessarily 
be deterministic: a suitably chosen probability measure on 
the space of sets might do the job.  

To solve the set selection problem would (almost certainly) need 
some mathematical characterisation of quasiclassicality --- 
the combination of sporadic quantum unpredictability and 
generally deterministic classical evolution, following 
simple of equations of motion, that characterises our 
physical world.  
Attempts have been made to find such a characterisation 
by refining the consistent histories formalism.\refs{\gmhprd,\gmhstrong,
\kentmcelwaine} The problems encountered
seem formidable, and it is hard to believe a
general solution will be found in the foreseeable future. 
Perhaps there is none. 

It is sometimes necessary to step back in 
order to make progress.  It seems at least worth 
considering the possibility that the criterion 
of consistency is too restrictive, and that the set selection
problem should be addressed within the broader quantum histories
framework.  In fact, as the next section explains, 
in the non-relativistic case at least, much more 
progress can be made this way: dynamical collapse models can be 
naturally reinterpreted as candidate solutions to the set 
selection problem.

\newsec{Unification of quantum history and dynamical reduction approaches} 

Ghirardi-Rimini-Weber's ``spontaneous localisation'' or 
``quantum jump'' model,\refs{\grw} 
lucidly explained in simple terms by Bell,\refs{\bellgrw}
is the ur-model of modern dynamical collapse theories. 
In an appropriate limit, it leads to one of a class of
Markovian stochastic differential equations,\refs{\gpr} which 
define testable alternatives to the Schr\"odinger equation. 
Several concrete proposals of this 
type\refs{e.g.\  \gpr, \ggr, \gisin, \diosi} have been put forward,
as well as more speculative ideas concerning possible relativistic 
generalisations.\refs{e.g.\  \pearle, \percival}
While this has undoubtedly been a very fruitful 
direction to pursue, it is surely not the only 
interesting way of extending the original GRW model.
I would like to suggest another path here. 

Recall that, according to the original GRW model, defined 
for $N$ distinguishable spinless particles, the wave function
\eqn\wfd{
\psi( x_1 , \ldots , x_N ; t ) }  
undergoes two types of evolution.  Almost all of the time, it follows
the Schr\"odinger equation, but at discrete randomly chosen times 
it jumps discontinuously, so that 
\eqn\jump{
\psi \rightarrow C \exp( -( x_i - x )^2 / 2 a^2 ) \psi \, , }
where particle $i$ is chosen randomly from $1$ to $N$, the 
coordinate $x$ is chosen randomly from the distribution
\eqn\xdistn{
\int d^3 x_1 \ldots d^3 x_N \exp (-( x_i - x )^2 /  a^2 ) | \psi |^2 
\, , }
$a$ is a constant parametrising the model, and $C$ is chosen so that
the new wave function is normalised.  
The times of these jumps are defined by a Poisson process, with
mean interval $\tau / N$ between jumps.  The parameters $\tau$ and $a$
are to be thought of here as new constants of nature; GRW originally
suggested
\eqn\params{ a \approx 10^{-5} {\rm~cm} , \qquad 
 \tau \approx 10^{15} {\rm~sec}.}

Of course, this is rather ad hoc, and no one seriously believes
that these equations --- or any of the models proposed to date ---
are likely to be precisely correct.    
But the GRW model and its successors demonstrate that 
mathematically precise theories can be found from which both 
the Schr\"odinger equation and the projection postulate can 
be derived as approximations,\refs{\grw, \bellgrw} in a way which extends 
to indistinguishable particles,\refs{\gpr} 
and with parameters that can be chosen consistent with 
experiment, so far as is known.\refs{\pearlesquires}

To rephrase the model in the language of histories, note that
the jump equation \jump\ corresponds to an unsharp event 
defined by a positive operator $A^i_x$ whose action on 
wave functions $\phi$ is 
\eqn\povai{ A^i_x : \phi \rightarrow \exp( - ( x_i - x )^2 /  a^2 ) \phi
  \, ,} 
so that up to normalisation \jump\ can be written as 
\eqn\jumpre{
\psi \rightarrow ( A^i_x )^{\half} \psi \, . }
The operators $A^i_x$ define a continuous decomposition of 
the identity:
\eqn\decompai{ {1 \over N \sqrt{a \pi}} \sum_i \int d^3 x A^i_x  = I
  \, .}
The probability distribution \xdistn\ can equivalently be written as  
\eqn\pdrewrite{
\langle \psi | A^i_x  | \psi \rangle =  
\Tr( (A^i_x )^{\half} | \psi \rangle \langle \psi |  (A^i_x
)^{\half} ) \, .}
(This is no accident: GRW's definitions 
were motivated by the theory of unsharp measurements.) 

Now, if the initial
state $\rho = | \psi \rangle \langle \psi |$, then 
the probability \povprob\ for the
quantum history defined by a series of
unsharp events chosen from decompositions \decompai\ is 
\eqn\provprobre{
\Tr ( (A^{i_n}_{x_n} )^{\half} \ldots 
(A^{i_1}_{x_1} )^{\half} \rho ( A^{i_1}_{x_1} )^{\half} \ldots 
(A^{i_n}_{x_n} )^{\half} ) \, .}
Translating from the Heisenberg picture to the Schr\"odinger, 
we see that \jump\ and \xdistn\ are precisely the outcomes and 
probabilities for unsharp events of this kind.  

Rewritten in this way, the GRW model defines 
a probabilistic set selection rule for a quantum histories
formulation based on unsharp events.
The set is selected by the choice of decompositions \decompai\ together
with the random choice of Poisson times; its 
histories are given by sequences of unsharp events $A^i_x$ 
at the chosen times.  
Since continuous stochastic equations of quite general form
can be arbitrarily well approximated\refs{\gpr} by discrete 
jump models of GRW type, the later dynamical collapse model
proposals\refs{\gpr, \ggr, \gisin, \diosi} can also 
effectively be interpreted in the same way.  
The selected sets, however, violate \sumrulespov\ and so 
are not consistent --- which is why the models disagree 
with standard quantum theory.  
If one is willing to pay this price (without necessarily 
contradicting experiment), quasiclassicality is not so 
hard to characterise.

\newsec{Conclusions and prospects} 

Things happened in the past, which were unobserved at the time, 
and whose consequences it is now  
impractical to describe in terms of present 
observations; to understand the present state
of the world properly, we need to be able to include  
such past events in our theories --- these may not be 
unquestionable assumptions, but they do not seem particularly 
outlandish.  They could 
turn out to be more or less forced on
us by accumulating cosmological data.  

In any case, it seems worth trying to incorporate
them into quantum theory.  The least radical way of 
doing so is to try to find a natural representation 
of past events in some standard approach to quantum 
theory --- perhaps as projections, positive operators, 
or partitions of the path integral --- and then to 
try to define some probabilistic interpretation 
in which histories of events are the primary objects.
This leads naturally to some form of quantum histories
approach, in which histories are grouped into complete
sets of exclusive alternatives, on each of which sets
a probability measure is defined.  

One can then, by   
trying to characterise interesting mathematical 
properties of sets of histories, try to develop criteria 
which select out sets that might 
be particularly physically interesting.  
Any criterion defines a new quantum 
histories formalism; all of these formalisms have a 
natural interpretation.  
This is the route pioneered by Griffiths,\refs{\griff} Omnes,\refs{\omnes}
Gell-Mann and Hartle,\refs{\gmhsantafe} who have set out a 
consistent (or decoherent) histories interpretation of quantum 
theory based on particular choices of criteria; 
stronger\refs{\akordered} and 
weaker\refs{\goldsteinpage,\pitowskyhemmo} natural criteria 
have also been found. 

As a research program, the quantum histories approach has been, 
and presumably will continue to be, very productive, raising many 
new and interesting questions.
However, considered as a finished product, the consistent (or
decoherent) histories interpretation must, I believe, be judged 
a failure as a scientific theory.
Its relation to classical mechanics and Copenhagen quantum mechanics
is very different from, for example, that of general relativity to 
Newtonian gravity and fluid dynamics.
And the comparison is not to its advantage: it is unable to 
account for the simplest predictions
or retrodictions, or to explain the success of Copenhagen quantum
mechanics or classical mechanics.\refs{\dowkerkentone, \dowkerkenttwo}

Even judged as mathematical criteria, consistency and medium decoherence,
while undoubtedly interesting properties of sets of histories, 
involve arbitrary choices and have some decidedly unnatural 
features.\refs{\akcontrary}  
It may, in any event, be more sensible to treat the existing
criteria as useful taxonomic labels rather than 
as badges of validity.  
Perhaps some inconsistent sets of histories will turn out 
to be scientifically useful; certainly almost all consistent 
sets will not.  

The key scientific problem in quantum histories approaches is to 
find some set selection rule, probabilistic or deterministic, 
sufficiently strong that it allows classical mechanics, Copenhagen 
quantum mechanics, and quantum field theory to be derived within
characterisable domains of validity.  
(Given a quantum theory of gravity, one would similarly hope to 
be able to derive general relativity.)  

It is an open question whether any precise rule of this type can be found 
within the consistent histories approach, even in the non-relativistic
case.  The attempts to date do not inspire overwhelming optimism. 
However, by going outside the consistent histories framework, and 
deviating from standard quantum mechanics, a solution to the non-relativistic
set selection problem can be found, by reinterpreting dynamical
collapse models of Ghirardi-Rimini-Weber type in the 
framework of quantum histories.  

Encouragingly from the point
of view of relativistic generalisation, the quantum histories 
framework includes covariantly defined notions of 
event.\refs{\hartleone} In this sense each approach seems to 
hold out the prospect of a solution to the deepest problem of the other.  
A covariantly defined set selection rule, which picks out generally
inconsistent sets and reduces to something resembling a dynamical
collapse model in the non-relativistic limit, would be a particularly
attractive way of solving the deep problem of interpreting quantum
theory in the cosmological context, since it need not necessarily
require any great conceptual revolution that threatens the 
successes of our present theories or (most of) their fundamental 
principles.  It would, of course, disagree at least subtly with 
the predictions of standard quantum theory --- but then, if nature 
really has chosen to make fundamental use of the notion of a quantum event, 
it would seem uncharacteristically tasteless to have done so in a 
way that leaves such events entirely undetectable.   

Though the line of thought which leads to this last speculative
proposal could, 
of course, be wrong in any of several places, it seems to 
me to strengthen the case for taking dynamical collapse models 
seriously.  

\vskip15pt
\leftline{\bf Acknowledgments}

I am very grateful to Fay Dowker and Jim McElwaine for 
collaborations reported here and many invaluable discussions,
to Oliver Rudolph for a helpful correspondence on unsharp events,
and to Charlotte Bonardi and Patrick Rabbitt for kindly 
clarifying the history of behaviourism for me. 
This work was supported by a Royal Society University Research
Fellowship.  
 
\listrefs
\end